\documentclass[prb,twocolumn,epsfig,showpacs]{revtex4}

\usepackage{graphicx}
\usepackage{dcolumn}
\usepackage{bm}
\usepackage{epsfig}

\begin{document}

\title{Long wavelength spatial oscillations of high frequency current noise in 1D electron systems}
\author{B.~Trauzettel}
\email{tzettel@physik.uni-freiburg.de}
\affiliation{Physikalisches Institut, Albert-Ludwigs-Universit\"at, D-79104
Freiburg, Germany}
\author{H.~Grabert}
\affiliation{Physikalisches Institut, Albert-Ludwigs-Universit\"at, D-79104
Freiburg, Germany}

\date{\today}

\begin{abstract}
Finite frequency current noise is studied theoretically for a
1D electron system in presence of a scatterer. In contrast to zero frequency
shot noise, finite frequency noise 
shows spatial oscillations at high frequencies with wavelength $\pi v_F/\omega$. Band curvature leads to
a decay of the amplitude of the noise oscillations as one moves
away from the scatterer, superimposed by a beat. Furthermore, Coulomb
interaction reduces the amplitude and modifies the wavelength of the
oscillations, which we inspect in the framework of the Luttinger liquid (LL)
model. The oscillatory noise contributions are only suppressed altogether when the LL interaction parameter $g \rightarrow 0$.
\end{abstract}


\maketitle

\section{Introduction}
\label{intro}
Current noise in mesoscopic systems has become an attractive research area,
because it contains information that cannot be extracted from the
current-voltage characteristics alone \cite{blant00}. Most of the
theoretical discussions about current noise in one-dimensional (1D) systems
focus on the zero frequency shot noise, because it may provide insights into
quantization of charge and statistics of charge carriers in a quantum
wire. In earlier theoretical work non interacting (Fermi liquid) 1D systems were analyzed in
the spirit of the Landauer-B\"uttiker scattering matrix formalism
\cite{lesov89,butti90,butti92,marti92}. Short wavelength spatial oscillations
of shot noise close to an impurity in the quantum wire
were investigated by Gramespacher and B\"uttiker \cite{grame99} based on the
assumption of non interacting charge carriers. More recently, shot noise of
strongly correlated electron systems has been addressed using
different algebraic techniques, such as for instance bosonization
\cite{kane94}, the thermodynamic Bethe ansatz \cite{fendl95a}, and
refermionization \cite{chamo96}. It has been shown by\ different groups \cite{ponom99a,trauz02} that due to the coupling of a
strongly correlated (Luttinger liquid) electron system to noninteracting
(Fermi liquid) electron reservoirs, the ideal Poisson shot noise is not renormalized by the interaction, as might have been expected from
earlier theoretical work \cite{kane94,fendl95a,chamo96}.

Most of the theoretical investigations on current noise calculate the finite
frequency noise at the impurity site, so that spatial oscillations remain
unnoticed. However, Lesovik \cite{lesov99} has calculated the current-current correlator
$\langle I(t,x) I(t',x') \rangle$, and therefore the integrand of the current
noise given by Eq.~(\ref{cndef}) below, in the most general form ($t \neq t'$,
$x \neq x'$), but he has only analyzed this correlator for the case of
a clean quantum wire, where spatial oscillations of finite frequency
current noise do not exist. How averaging over space affects the shot noise of
a two- and three-terminal 1D system was investigated by Gavish, Levinson, and
Imry \cite{gavis01}. Furthermore, the wave behavior of a high-frequency current
through a double-barrier tunneling structure was studied numerically by
Cai, Hu, and Lax \cite{cai91}. Although most of these works have investigated
certain aspects of the spatial structure of the current noise, a discussion of
the long wavelength spatial dependence of finite frequency noise is
lacking. This will be done in this article. 

The spatial dependence of the noise becomes important in the high frequency
regime, because, apart from the length of the quantum wire, $v_F/\omega$, where $\omega$ is the noise frequency and $v_F$
the Fermi velocity, provides an additional length scale, which affects current
noise measurements. Only for $\omega \rightarrow 0$ charge conservation
implies a noise strength independent of the position, where it is measured. We
will show that once $\omega$ is finite, the current noise exhibits long wavelength spatial oscillations, which
depend on the noise frequency, the Fermi velocity, and the distance between the
point of measurement and the impurity in the quantum wire. We first address
non interacting systems with linear dispersion, where the origin of the
spatial oscillations is easily seen. However, these
noise oscillations are affected by band curvature and Coulomb interaction. A
finite band curvature calculation shows that the spatial oscillations decay away from the impurity site and show a beating behavior. Coulomb
interaction, which we take into account by describing the quantum wire as a
Luttinger liquid (LL), suppresses the amplitude of the noise oscillations by a
factor $g$ and modifies its wavelength. Here, $g$ is the LL
interaction parameter ($g<1$ for repulsive interactions).

The article is organized as follows. In Chapter \ref{theo} we describe the
general model of a quantum wire with an impurity. Then, in Chapter
\ref{cunoise} we calculate the finite frequency noise with special emphasis on
the noise oscillations. Chapter \ref{curv} treats the influence of band
curvature, and Chapter \ref{coul} the effect of Coulomb interaction on the
spatial oscillations. Finally, we summarize our results and point out open
problems in Chapter \ref{conclu}. Throughout this article we formally treat
spin-polarized (single channel) quantum wires, but our findings
apply generally to the case of electrons with spin. For non interacting
electrons, the effect of spin on the noise is trivial, leading to an overall
factor 2. The results in Chapter \ref{coul} for electrons with Coulomb
interaction, are also only trivially affected by the inclusion of spin, again
leading to an overall factor 2. This is due to the fact that we restrict the
discussion in Chapter \ref{coul} to the spatial oscillations of the
equilibrium noise in a half-open wire, where charge and spin field operators
are decoupled \cite{halda81}.
  
\section{General model}
\label{theo}

Non interacting electrons of mass $m$ moving along a quantum wire with a
$\delta$-scatterer at $x=0$ are described by the Schr\"odinger equation
\begin{equation} \label{schroed}
\left( -\frac{\hbar^2}{2m} \frac{\partial^2}{\partial x^2} +
\frac{\hbar^2}{m}\Lambda \delta(x) \right) \psi(x) = E \psi(x)
\; ,
\end{equation}
where the scattering potential is $V(x)=(\hbar^2 \Lambda/m)\delta(x)$ with a
scattering strength $\Lambda$. For all positive energies $E_k=\hbar^2
k^2/2m$, the eigenstates ($k>0$)
\begin{equation} \label{rmo}
\psi_k (x) = \frac{1}{\sqrt{L}} \left\{ \begin{array}{r@{\quad \quad}l} e^{ikx}
+ r_k \ e^{-ikx} & (x<0) \\ t_k \ e^{ikx} & (x>0) \end{array} \right. \; ,
\end{equation}
where $L$ is the length of the quantum wire, describe a wave incident from the
left, commonly called {\em right-mover}, that is partially transmitted and partially reflected and solves Eq.~(\ref{schroed}) with a transmission amplitude
\begin{equation} \label{tk}
t_k = \frac{1}{1+i\Lambda/k} \; .
\end{equation}
Likewise, there is a solution ($k>0$)
\[
\phi_k (x) \equiv \psi_{-k}(x) = \psi_k(-x)
\]
describing a wave incident from the right, a so called {\em left-mover}
eigenstate. We apply a voltage $U$ to the ends of the quantum wire by
connecting them adiabatically to external leads with chemical potentials
$\mu_L=E_F+eU$ and $\mu_R=E_F$, where $E_F$ is the Fermi energy. Consequently,
the states $\psi_k (x)$ are filled up to $k=(k_F^2+2meU/\hbar^2)^{1/2}$, while
the states $\phi_k(x)$ are only filled up to $k=k_F$. If we assume a rather
small applied voltage $U$ (compared to $E_F$), the transmission ($t_k$) and
reflection amplitudes ($r_k$) do not vary much in the $k$-range $[k_F,(k_F^2+2meU/\hbar^2)^{1/2}]$ and can effectively be replaced by constant values at the Fermi edge, $t_{k_F}\equiv t$ and $r_{k_F} \equiv r$.
 
In the second quantized form, the eigenstates (\ref{rmo}) allow to expand the electron field operator
\begin{equation} \label{elecop}
\Psi(t,x) = \sum_{k>0} \left( a_k \psi_k(x) + b_k \phi_k(x) \right) e^{-(i/\hbar) E_k t} \,
\end{equation}
where the operators $a_k^\dagger$ ($b_k^\dagger$) and $a_k$ ($b_k$) are Fermi
creation and annihilation operators obeying the usual anticommutation
relations. Then, in the grand canonical ensemble
\[
\rho = \frac{e^{- \beta \sum_k \left[ \left( E_k-\mu_L \right) a^\dagger_k
a^{}_k + \left( E_k-\mu_R \right) b^\dagger_k b^{}_k \right] }}{ {\rm tr}
\left\{ e^{- \beta \sum_k \left[ \left( E_k-\mu_L \right) a^\dagger_k
a^{}_k + \left( E_k-\mu_R \right) b^\dagger_k b^{}_k \right] }\right\}}
\]
the expectation values of products of creation and annihilation operators read
\begin{eqnarray*}
\langle a_{k_1}^\dagger a^{}_{k_2} \rangle &=& f(k_1;U) \delta_{k_1 k_2} \; , \\
\langle a_{k_1}^\dagger b^{}_{k_2} \rangle &=& 0 \; , \\
\langle b_{k_1}^\dagger b^{}_{k_2} \rangle &=& f(k_1;0) \delta_{k_1 k_2}
\end{eqnarray*}
with the standard Fermi distribution function
\[
f(k;U) = \frac{1}{1+\exp[\beta(E_k-(E_F+eU))]} \; ,
\]
where $\beta^{-1} = k_B T$. For future reference, we now briefly discuss the calculation of the current
through a 1D non interacting quantum wire. Inserting the expansion (\ref{elecop}) into the standard expression for the quantum mechanical current operator $I(t,x)$ yields
\begin{eqnarray} \label{iop}
&& I(t,x) = \frac{e}{2mi} \sum_{k,k'} e^{(i/\hbar)(E_k-E_k')t} \times
\nonumber \\
&& \Bigl(a_k^\dagger \psi^*_k +b_k^\dagger \phi_k^* \Bigr) \hat{\nabla}
\Bigl(a_{k'} \psi_{k'} +b_{k'} \phi_{k'} \Bigr) \;  ,
\end{eqnarray}
where the operator $\hat{\nabla}$ is defined by $A\hat{\nabla}B=A(\nabla
B)-(\nabla A)B$. Note, that in Eq.~(\ref{iop}) and in the rest of the article all $\psi_k$'s and $\phi_k$'s depend explicitly on $x$. Then, the expectation value of the current becomes
\begin{eqnarray} \label{iexp1}
&&\langle I \rangle = \frac{e}{2mi} \sum_k \times \nonumber \\
&& \Bigl( \langle a_k^\dagger a_k^{} \rangle \psi^*_k \hat{\nabla} \psi_k +
\langle b_k^\dagger b_k^{} \rangle \phi^*_k \hat{\nabla} \phi_k \Bigr) \; .
\end{eqnarray} 
If we now consider the quantum wire, say, on the right side of the impurity ($x>0$) we find
\begin{eqnarray*}
\psi_k^* \hat{\nabla} \psi_k &=& \frac{i2k}{L} |t|^2 = \frac{i2k}{L} {\cal T} \; , \\
\phi_k^* \hat{\nabla} \phi_k &=& - \frac{i2k}{L} (1-|r|^2) = -
\frac{i2k}{L} (1-{\cal R}) \; .
\end{eqnarray*}
Here, ${\cal T}=|t|^2$ is the transmission coefficient of the current flow through
the barrier and ${\cal R}=|r|^2=1-{\cal T}$ the reflection coefficient. Going
over to the
continuum limit ($L \rightarrow \infty$) and changing variables $k \rightarrow E_k = \hbar^2 k^2/2m$,
the current reads
\begin{eqnarray} \label{landauform}
\langle I \rangle &=& \frac{e^2}{h} {\cal T} \int dE_k \left( f(E_k;U)-f(E_k;0)
\right) \nonumber \\
&=& \frac{e^2}{h} {\cal T} U \; .
\end{eqnarray}
This is the well-known Landauer formula for the current through a quantum
wire. Without inelastic scattering in the wire,
Eq.~(\ref{landauform}) holds for finite, but moderate temperatures and applied
voltages ($k_B T, eU \ll E_F$) \cite{datta95}.

\section{Noise oscillations}
\label{cunoise}

The finite frequency current noise at position $x$ of the quantum wire is
defined by the expression
\begin{equation} \label{cndef}
P(\omega,x) = \int dt e^{i\omega t} \left\langle \left\{ \Delta I(t,x), \Delta
I(0,x) \right\}_+ \right\rangle \; ,
\end{equation}
where $\Delta I(t,x) = I(t,x) - \langle I \rangle$ is the current fluctuation
operator. Without loss of generality, we assume $\omega>0$ and
$x>0$. Inserting the current operator (\ref{iop}) into the definition of the
current noise (\ref{cndef}), we find
\begin{eqnarray} \label{sn2}
&& P(\omega,x) = \nonumber \\
&& - \frac{e^2}{4m\hbar L} \sum_{k_1, \dots, k_4} \frac{1}{|k_1|}
\delta_{k_1,(k_2^2-2m\omega/\hbar)^{1/2}} \times \nonumber \\
&& \Bigl\langle \Bigl[ \left(
a_{k_1}^\dagger \psi_{k_1}^* + b_{k_1}^\dagger \phi_{k_1}^* \right)
\hat{\nabla} \left( a_{k_2} \psi_{k_2} + b_{k_2} \phi_{k_2} \right) \Bigr]
\times \nonumber \\
&& \Bigl[ \left( a_{k_3}^\dagger \psi_{k_3}^* + b_{k_3}^\dagger
\phi_{k_3}^* \right) \hat{\nabla} \left( a_{k_4} \psi_{k_4} + b_{k_4}
\phi_{k_4} \right) \Bigr] \Bigr\rangle_c \nonumber \\
&& + \;\; {\mbox {h.c.}} \; ,
\end{eqnarray}
where the $t$-integration in Eq.~(\ref{cndef}) has already been
carried out, and the symbol $\langle \dots \rangle_c$ represents the connected
expectation value. Exploiting Wick's theorem, see e.g. Ref.~\onlinecite{mahan90}, we can
only pair $k_1=k_4=(k_2^2-2m\omega/\hbar)^{1/2}$ and $k_2=k_3$. To simplify our notation, we set $k_2=k$ and define $k_\omega \equiv (k^2-2m\omega/\hbar)^{1/2}$. Then, the resulting expression for the current noise reads
\begin{eqnarray} \label{sn3}
&& P(\omega,x) = - \frac{e^2}{4m\hbar} L \sum_{k>0} \frac{1}{|k_\omega|}
\times \nonumber \\
&& \Bigl\{ f^{aa}(k,\omega) \left(\psi_{k_\omega}^* \hat{\nabla} \psi_{k}
\right) \left(\psi_{k}^* \hat{\nabla} \psi_{k_\omega} \right) \nonumber \\
&& + f^{bb}(k,\omega) \left(\phi_{k_\omega}^* \hat{\nabla} \phi_{k} \right)
\left(\phi_{k}^* \hat{\nabla} \phi_{k_\omega} \right) \nonumber \\ 
&& + f^{ab}(k,\omega) \left(\psi_{k_\omega}^* \hat{\nabla} \phi_{k}
\right) \left(\phi_{k}^* \hat{\nabla} \psi_{k_\omega} \right)
\nonumber \\
&& + f^{ba}(k,\omega) \left(\phi_{k_\omega}^* \hat{\nabla} \psi_{k} \right)
\left(\psi_{k}^* \hat{\nabla} \phi_{k_\omega} \right) \Bigr\} \; , 
\end{eqnarray}
where we have introduced the short-hand notation
\begin{eqnarray*}
&& f^{\alpha \beta} (k,\omega) = \\
&& f^\alpha(k_\omega)(1-f^\beta(k)) + (1-f^\alpha(k_\omega))f^\beta(k)
\end{eqnarray*}
with $f^a=f(k;U)$ and $f^b=f(k;0)$. In the remaining part of this section we
assume a linear dispersion relation, which is justified for instance in single
wall carbon nanotubes \cite{lemay01} and is also appropriate to cleaved edge overgrowth quantum
wires \cite{ausla02} for small applied voltages $U$ and low temperatures $T$. In
that case, the energy $E_k$ of the right-mover eigenstates may be replaced by
$\hbar v_F (k-k_F)$ and analogously the energy of the left-mover eigenstates
by $-\hbar v_F (k-k_F)$. We then find for the finite frequency current noise
($x>0$) at zero temperature
\begin{eqnarray} \label{sn4}
&& P(\omega,x) = \frac{\hbar^2 e^2 v_F}{(2m)^2} \times \nonumber \\
&& \left\{ \int\limits_{eU/\hbar v_F+k_F^-}^{eU/\hbar v_F+k_F}
\frac{dk}{2\pi} \ \left( 2k+\frac{\omega}{v_F}\right)^2 {\cal T}^2
\right. \nonumber \\
&& + \int\limits_{k_F^-}^{eU/\hbar v_F+k_F} \frac{dk}{2\pi} \ {\cal
A}(k,\omega) \nonumber \\
&& + \int\limits_{k_F^-}^{k_F} \frac{dk}{2\pi} \left[ \left(
2k+\frac{\omega}{v_F} \right)^2 \right. \times \nonumber \\
&& \left. \Bigl( 1-2{\cal R} \cos\left( \frac{2 \omega
x}{v_F} \right) + {\cal R}^2 \Bigr) + 2 \left(\frac{\omega}{v_F} \right)^2
{\cal R} \right]  \nonumber \\
&& + \Theta(\hbar \omega-eU) \int\limits_{eU/\hbar v_F + k_F^-}^{k_F}
\frac{dk}{2\pi} \ {\cal A}(k,\omega) \nonumber \\
&& + \Theta(eU-\hbar \omega) \left. \int\limits_{k_F}^{eU/\hbar v_F+k_F^-}
\frac{dk}{2\pi} \ {\cal A}(k,\omega) \right\} ,
\end{eqnarray}
where $\Theta(x)$ is the Heaviside function, $k_F^- \equiv
k_F-\omega/v_F$, and
\[
{\cal A}(k,\omega) \equiv \left[
\left(2k+\frac{\omega}{v_F} \right)^2 {\cal R}{\cal T} +
\left(\frac{\omega}{v_F}\right)^2 {\cal T} \right] \; .
\]
Since we are interested in the current noise away from the
impurity site, in Eq.~(\ref{sn4}) we have omitted all contributions of
Eq.~(\ref{sn3}) containing factors of $\exp[i2kx]$, which decay like $1/x$
after doing the $k$-integrals in Eq.~(\ref{sn4}). The neglected
$\exp[i2kx]$-terms lead to short wavelength noise oscillations having the same origin as the well-known Friedel oscillations \cite{fried58}. If we now integrate over $k$ in Eq.~(\ref{sn4}) we obtain for $eU, \hbar \omega \ll E_F$ the current noise
\begin{eqnarray} \label{noiseresult}
&& P(\omega,x) = \nonumber \\
&& \frac{e^2}{h} \Bigl\{ \left( |eU+\hbar\omega| +
|eU-\hbar\omega| \right) {\cal R}{\cal T} \nonumber \\
&& + \hbar |\omega| \left( 1-2{\cal R}\cos\left(\frac{2\omega
x}{v_F}\right)+{\cal R}^2 + {\cal T}^2 \right) \Bigr\} \; . \nonumber \\
\end{eqnarray}
With $\cos (2\omega x/v_F) \rightarrow 1$, Eq.~(\ref{noiseresult}) was derived
earlier by Yang \cite{yang92}. However, this low frequency approximation is
not appropriate to determine the finite frequency current noise {\em far away}
from the impurity. Let us consider two limits of Eq.~(\ref{noiseresult}). For
$\omega=0$, the shot noise $P(0,x)=2(e^3/h){\cal RT} |U|$ is independent of
the point of measurement as it should be. On the other hand, the equilibrium
noise for $U=0$ reads
\begin{eqnarray*}
&& P_0(\omega,x) = \frac{e^2}{\pi} |\omega| \times \\
&& \left\{ {\cal RT} + \frac{1}{2}\left( 1-2{\cal R}\cos\left(\frac{2\omega
x}{v_F}\right)+{\cal R}^2 + {\cal T}^2 \right) \right\} \; .
\end{eqnarray*}
As a consequence of the fluctuation-dissipation theorem
\cite{butti93}, $P_0(\omega,x)$ is related to the real (dissipative) part of
the frequency-dependent conductance by
\begin{equation} \label{gprela}
P_0(\omega,x) = 2 \hbar |\omega| \; {\mbox{Re}} [G(\omega,x)] \; .
\end{equation} 
Therefore, ${\mbox{Re}} [G(\omega,x)]$ is an oscillating
function of frequency with a period $\pi v_F/x$ and shows peaks with a maximum
height of $2e^2/h$, only truly reached for ${\cal R} = 1$. These findings are
in agreement with the analysis by Blanter, Hekking, and B\"uttiker of the
dynamic conductance of an interacting quantum wire capacitively coupled to a gate \cite{blant98}.
 
The appearance of the $\cos(2\omega x/v_F)$-term in Eq.~(\ref{noiseresult}) is
the main result of this chapter. Note, that the noise oscillations with the
long wavelength $v_F/\pi\omega$ have to be distinguished from the {\em
Friedel-type} noise oscillations with the short wavelength $\pi/k_F$ discussed
in Ref.~[\onlinecite{grame99}]. Evidently, for a very strong backscatterer (${\cal
R} \approx 1$) the oscillating part of Eq.~(\ref{noiseresult}) is most
pronounced. Therefore, to further analyze the noise oscillations, we will
focus on the situation of a half-open quantum wire (${\cal R} = 1$ at the
boundary) in the following. Then, the shot noise contribution of
Eq.~(\ref{noiseresult}) vanishes and the equilibrium finite frequency current
noise is given by
\begin{equation} \label{pho}
P(\omega,x) = 2 \frac{e^2}{h} \hbar |\omega| \left[ 1-
\cos\left( \frac{2\omega x}{v_F} \right) \right] \; .
\end{equation}
At the point $x$ of measurement of the current noise the incoming and the
reflected {\em current noise waves} interfere, leading to
a suppression or an enhancement of the measured noise depending on the time
$2x/v_F$ an electron needs to return to the position $x$ after a reflection at
$x=0$. Remarkably, the noise oscillations at zero temperature in
Eq.~(\ref{pho}) do not decay away from the open boundary -- a situation
certainly due to idealizations. To see if and how those noise oscillations
survive in more realistic systems we explicitly treat finite band curvature and
intrinsic electron-electron interaction in the next two chapters.

\section{Effect of band curvature}
\label{curv}

\begin{figure}
\vspace{0.5cm}
\begin{center}
\epsfig{file=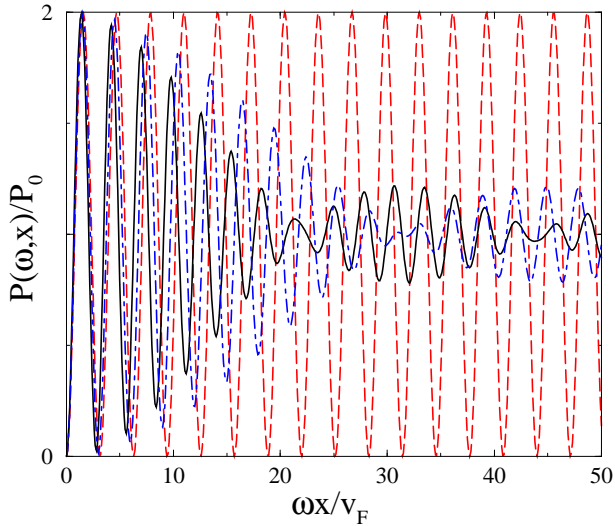,scale=0.5}  
\caption{\label{compare} Finite frequency current noise $P(\omega,x)$
in units of $P_0 = 2 (e^2/h) \hbar |\omega|$ as a function of $\omega x/v_F$
for a half-open quantum wire. The (black) solid line shows the result
(\ref{pcurv}) corresponding to a 1D electron system with a quadratic
dispersion relation, the (blue) dot-dashed line illustrates the approximate
result (\ref{pexp}) for $E_0 \equiv \hbar \omega/E_F = 0.1$,
and the (red) dashed line corresponds to Eq.~(\ref{pho}), the current noise of
an electron system with linear dispersion. It is clearly visible that a finite
band curvature modifies the amplitude and the phase of the noise oscillations.}
\end{center}
\end{figure}

Band curvature of the dispersion relation of electrons induces a decay of the noise oscillations
away from the scatterer. To investigate how fast these noise oscillations decay, we go back to Eq.~(\ref{sn3}), but keep only track of the non-vanishing
terms for a half-open quantum wire, where ${\cal R}=1$ (${\cal
T}=0$). Again we neglect all ($2k_F x$)-oscillating contributions. Then, we
find the following expression for the current noise at zero temperature
\begin{eqnarray} \label{pcurv}
&& P(\omega,x) = \nonumber \\
&& 2 \frac{\hbar e^2}{m}
\int\limits^{k_F}_{k_F^\omega} \frac{dk}{2\pi} k \Bigl[ 1-\cos\left(
2\left( k_\omega - k \right) x \right) \Bigr]
\end{eqnarray}
with $k_F^\omega=(k_F^2-2m\omega/\hbar)^{1/2}$ and $k_\omega=(k^2-2m\omega/\hbar)^{1/2}$. While the result for $P(\omega,x)/P_0$ (with
$P_0 = 2 (e^2/h) \hbar |\omega|$) derived in the previous section, see Eq.~(\ref{pho}), only depends on a single
dimensionless parameter, namely $\omega x/v_F$, in Eq.~(\ref{pcurv}) a second
dimensionless parameter arises. This is $E_0 \equiv \hbar
\omega/E_F$, where $E_F=\hbar^2 k_F^2/2m$. In the limit $E_0 \rightarrow
0$ with $\omega x/v_F$ fixed, the full result (\ref{pcurv}) reduces to
Eq.~(\ref{pho}), the result with linear dispersion. A simple expansion of
Eq.~(\ref{pcurv}) in powers of $E_0$ does not capture the qualitive behavior
of the full solution (\ref{pcurv}), characterized by a decay and a beat of the
noise oscillations as a function of $\omega x/v_F$ (see Fig.~\ref{compare}). Similarly, an asymptotic
expansion for large $\omega x/v_F$, which could explain the decay, does not
give any information about the interesting parameter regime $\omega x/v_F \in
[0,2\pi]$. To further analyze the behavior of Eq.~(\ref{pcurv}), we have
expanded the lower boundary $k_F^\omega$ as well as $k_\omega$ in the
integrand in powers of $2m\omega/\hbar k_F^2$ and $2m\omega/\hbar k^2$,
respectively. This procedure is reasonable for $\hbar \omega \ll E_F$,
because the integration variable $k$ only takes values close to $k_F$,
$k \in [k_F^\omega,k_F]$, and the integrand of Eq.~(\ref{pcurv}) is smooth for
$k$ near $k_F$. Then, we can do the integration in Eq.~(\ref{pcurv}) easily and find
\begin{eqnarray} \label{pexp}
&& \frac{P(\omega,x)}{P_0} = \nonumber \\
&& 1- \frac{2v_F \sin\left(\frac{E_0 \omega x}{2v_F}\right)}{E_0 \omega x}
\cos\left(\frac{2\omega x}{v_F} - \frac{E_0 \omega x}{2v_F}\right) \; .
\end{eqnarray}
Evidently, Eqs.~(\ref{pcurv}) and (\ref{pexp}) both reduce to Eq.~(\ref{pho})
in the limit $E_0 \rightarrow 0$ with $\omega x/v_F$ fixed. Furthermore,
Eqs.~(\ref{pcurv}) and (\ref{pexp}) show qualitatively the same oscillating
behavior superimposed by a beat as a function of $\omega x/v_F$. This is illustrated in Fig.~\ref{compare}, where the noise deduced from
Eqs.~(\ref{pho}), (\ref{pcurv}), and (\ref{pexp}) is plotted versus $\omega
x/v_F$.
A comparison of the full solution (\ref{pcurv}) and our simple model
(\ref{pexp}) tells us that the amplitude of the oscillations of $P(\omega,x)$
decays at least like $v_F/E_0 \omega x$ away from the scatterer and reaches
asymptotically the value $P_0$. Additionally, we also see that the phase of
the noise oscillations is modified by band curvature effects.

\section{Effect of Coulomb interaction}
\label{coul}

Now, we turn to the effect of Coulomb interaction on the noise oscillations. A
half-open strongly interacting 1D electron system can most easily be
described in the framework of open boundary bosonization
\cite{fabri95}. Following closely the notation of Fabrizio and Gogolin
\cite{fabri95} we set $\hbar \equiv 1$ in this section, except for the final result. Then, the full Hamiltonian of the system reads
\begin{equation} \label{obhfull}
H=H_0+H_I
\end{equation}
with the kinetic contribution
\begin{equation} \label{obh0}
H_0= \int_{-L}^0 dx \psi^\dagger (x)
\epsilon(-i\partial_x) \psi(x)^{}
\end{equation}
and the electron-electron interaction contribution
\begin{equation} \label{obhi}
H_I = \frac{1}{2} \int dx dy \  \psi^\dagger (x)
\psi^\dagger (y) U (x-y) \psi^{} (y) \psi^{} (x) \; ,
\end{equation}
where $\epsilon(k)$ is the dispersion relation of the 1D band, and $\psi(x)$
is the electron annihilation operator obeying the open boundary conditions
\begin{equation} \label{obc}
\psi(-L) = \psi(0) = 0 \; .
\end{equation}
To simplify the full Hamiltonian (\ref{obhfull}), Fabrizio and Gogolin have
bosonized the fermionic operators in Eqs.~(\ref{obh0}) and (\ref{obhi})
assuming a linear dispersion relation. The $\psi$ operator can be
written down in terms of right and left moving fields
\[
\psi(x) = e^{i k_F x}\psi_R(x) +e^{-i k_F x}\psi_L(x) \; .
\]
Due to the boundary conditions (\ref{obc}) these fields are not independent,
but satisfy by their definition \cite{fabri95}
\[
\psi_L(x) = - \psi_R(-x) \; .
\]
Therefore, all electron operators in the framework of open boundary
bosonization can be expressed in terms of e.g. right moving fields only, living
in the parameter space $x \in [-L,L]$. In the usual way, the
interacting electron field $\psi_R(x,t)$ can be bosonized with the help of a free boson field $\phi(x,t)$
\begin{equation}
\psi_R(x,t) = \frac{1}{\sqrt{2\pi a_0}} e^{i\phi(x,t)/\sqrt{g}} \; .
\end{equation} 
Here, $a_0$ is the lattice constant of the underlying lattice model and $g$ is
the LL interaction parameter, which is related to the Fourier transform
$\tilde{U}(k \rightarrow 0)$ of the interaction potential $U(x-y)$ in
Eq.~(\ref{obhi}) \cite{halda81}. Then, $0<g<1$ for repulsive interactions and
$g \rightarrow 1$ in the noninteracting limit. In the next step, the field
$\phi(x,t)$ may be expanded in terms of boson creation $b_q^\dagger$ and
annihilation $b_q$ operators, yielding
\begin{equation} \label{expan}
\phi_R(x,t) = \sum_{q>0} \sqrt{\frac{\pi}{qL}} \left[e^{iq(x-vt)} b_q +
e^{-iq(x-vt)} b_q^\dagger \right] e^{-a_0q/2} \; .
\end{equation} 
In that representation the momentum of the system takes quantized
values $q=\pi n/L$. With the approximations of linear dispersion and short
range interactions made, the bosonic Hamiltonian of the full interacting
fermion systems, Eq.~(\ref{obhfull}), takes a very simple form \cite{fabri95}
\begin{equation} \label{hamb}
H_b = v \sum_{q>0} q b_q^\dagger b_q \; ,
\end{equation}
where $v=v_F/g$ is the renormalized sound velocity.

In terms of the bosonic field the electron density of right- and left-moving excitations reads
\[
\rho_{R/L} (x,t) = \frac{k_F}{\pi} + \frac{1}{2\pi \sqrt{g}} \partial_x
\phi(\pm x,t) \; .
\] 
This allows us to write a bosonic form of the current operator
\begin{equation} \label{obcurrent}
I(x,t) = \frac{ev_F}{2\pi\sqrt{g}} \left( \partial_x \phi(x,t) - \partial_x
\phi(-x,t) \right) \; ,
\end{equation}
which we combine with Eq.~(\ref{cndef}) to compute the current noise
\begin{eqnarray} \label{bosnoise}
&& P(\omega,x) =  \left(\frac{ev_F}{2\pi\sqrt{g}}\right)^2 \int dt e^{i\omega
t} \times \\
&& \partial_x \left\langle \left\{ \phi(x,t) - \phi(-x,t),
\phi(x,0) - \phi(-x,0) \right\}_+ \right\rangle
\nonumber \; .
\end{eqnarray}
To proceed, we need to calculate the bosonic correlation function with respect
to the groundstate of the Hamiltonian (\ref{hamb}). At zero temperature, a straightforward calculation for large system size $L$ yields
\begin{equation} \label{phiphi}
\left\langle \phi(x,t) \phi(x',0) \right\rangle = - \ln \left(-i\frac{\pi}{L}
(x-x'-vt + i a_0) \right) \; .
\end{equation}
In the limit $L \rightarrow \infty$ this may be inserted into the expression
(\ref{bosnoise}) for the current noise to give
\begin{eqnarray} \label{intrep}
&& P(\omega,x) = -\frac{g e^2}{2\pi^2} \int dt \ e^{i
\omega t} \left( \frac{2}{(t-i a_0/v)^2} - \right. \nonumber \\
&& \left. \frac{1}{(2x/v - t + i a_0/v)^2} - \frac{1}{(2x/v + t - i
a_0/v)^2} \right) \; .
\end{eqnarray}
Doing the integral and then sending $a_0 \rightarrow 0$, the final result reads
\begin{equation} \label{obpfin}
P(\omega,x) = 2 g \frac{e^2}{h} \hbar |\omega| \left(1-\cos \left(\frac{
2\omega g |x|}{v_F} \right) \right) \; ,
\end{equation}
where we have re-introduced $\hbar$. A comparison of Eqs.~(\ref{pho}),
(\ref{pexp}), and (\ref{obpfin}) shows, that unlike band curvature
short range electron-electron interaction with a linear dispersion does not
lead to a decay of the noise oscillations, but instead suppresses the
amplitude by a factor $g$ and increases the wavelength of the oscillations by a
factor $1/g$. This shows that the noise oscillations are also present in an
interacting quantum wire with finite $g$. Only for very strong interactions
($g \rightarrow 0$) the noise oscillations will vanish. That the noise
oscillations survive in the presence of Coulomb interaction can be
understood as a consequence of the lack of strict electroneutrality in
a LL. In the presence of gates that screen the long range part of the
Coulomb interaction, only the part $(1-g^2) Q$ of a charge $Q$ on the
quantum wire is screened internally, the remaining part $g^2 Q$ being
compensated by the nearby gates \cite{eggra97}. Therefore, if $g>0$, which is
the typical experimental situation, the quantum wire can become charged and
exhibit the spatial noise oscillations (\ref{obpfin}).

\section{Conclusions}
\label{conclu}

To summarize, first, we have studied the finite frequency current noise of a
non interacting quantum wire with an impurity. In the equilibrium part of the
finite frequency current noise, which is related to the real part of the
dynamic conductance by means of the fluctuation-dissipation theorem, we have found
long wavelength spatial oscillations, which can be explained by simple quantum
mechanical wave dynamics. Those oscillations are absent in the shot noise
limit ($\omega \rightarrow 0$) as well as the ballistic limit (${\cal T}
\rightarrow 1$). The oscillations are most pronounced for a
half-open quantum wire, where the boundary can be regarded as a very
strong backscatterer. We have shown that finite band curvature leads to a moderate decay of the noise oscillations. The
strength of this decay is proportional to $E_0 \equiv \hbar \omega/E_F$, where
small $E_0$ means weak decay. Furthermore, we have discussed the noise
oscillations in strongly interacting half-open 1D systems using the LL
model. We have found that the amplitude of the spatial oscillations
decreases with increasing electron-electron interaction, while the wavelength
of the oscillations is enhanced by the LL interaction
parameter. Finally, we remark that all results for the current noise
derived here were obtained in the limit $L \rightarrow \infty$ of a very long
quantum wire and thus only hold for frequencies $\omega$ above $v_F/L$. To
treat frequencies below $v_F/L$, a finite $L$ calculation has to be carried
out. Then, in particular for the case of interacting electrons in the quantum
wire, the (non interacting) external reservoirs must explicitly be taken into account \cite{maslov}.

\begin{acknowledgements}
We wish to thank M.~B\"uttiker, M.~Devoret, R.~Egger, U.~Gavish, Y.~Imry, and
A.~Komnik for valuable discussions. This work has been supported by the
Deutsche Forschungsgemeinschaft under Grant No.~GR 638/19. 
\end{acknowledgements}


\begin{thebibliography}{10}

\bibitem{blant00}
{Ya.~M.~Blanter and M.~B\"uttiker}, Phys. Rep. {\bf {336}}, 1 (2000).

\bibitem{lesov89}
{G.~B.~Lesovik}, JETP Lett. {\bf {49}}, 592 (1989).

\bibitem{butti90}
{M.~B\"uttiker}, Phys.~Rev.~Lett. {\bf {65}}, 2901 (1990).

\bibitem{butti92}
{M.~B\"uttiker}, Phys.~Rev.~B {\bf {45}}, 3807 (1992).

\bibitem{marti92}
{T.~Martin and R.~Landauer}, Phys.~Rev.~B {\bf {45}}, 1742 (1992).

\bibitem{grame99}
{T.~Gramespacher and M.~B\"uttiker}, Phys.~Rev.~Lett. {\bf 81}, 2763 (1998);
Phys.~Rev.~B {\bf {60}}, 2375 (1999).

\bibitem{kane94}
{C.~L.~Kane and M.~P.~A.~Fisher}, Phys.~Rev.~Lett. {\bf {72}}, 724 (1994).

\bibitem{fendl95a}
{P.~Fendley, A.~W.~W.~Ludwig, and H.~Saleur}, Phys.~Rev.~Lett. {\bf {75}},
 2196 (1995).

\bibitem{chamo96}
{C.~de C.~Chamon, D.~E.~Freed, and X.~G.~Wen}, Phys.~Rev.~B {\bf {53}}, 4033
 (1996).

\bibitem{ponom99a}
{V.~V.~Ponomarenko and N.~Nagaosa}, Phys.~Rev.~B {\bf {60}}, 16865 (1999).

\bibitem{trauz02}
{B.~Trauzettel, R.~Egger, and H.~Grabert}, Phys.~Rev.~Lett. {\bf 88}, 116401
 (2002).

\bibitem{lesov99}
{G.~B.~Lesovik}, JETP Lett. {\bf {70}}, 208 (1999).

\bibitem{gavis01}
{U.~Gavish, Y.~Levinson, and Y.~Imry}, Phys.~Rev.~Lett. {\bf 87}, 216807
 (2001).

\bibitem{cai91}
{W.~Cai, P.~Hu, and M.~Lax}, Phys.~Rev.~B {\bf {44}}, 3336 (1991).

\bibitem{halda81}
{F.~D.~M.~Haldane}, J.~Phys.~C {\bf {14}}, 2585 (1981).

\bibitem{datta95}
{S.~Datta}, {\em Electronic Transport in Mesoscopic Systems} (Cambridge
  University Press, Cambridge, England, 1995).

\bibitem{mahan90}
{G.~D.~Mahan}, {\em Many-Particle Physics} (Plenum, New York, 1990).

\bibitem{lemay01}
{S.~G.~Lemay {\em {et al.}}}, Nature {\bf {412}}, 617 (2001).

\bibitem{ausla02}
{O.~M.~Auslaender {\em {et al.}}}, Science {\bf {295}}, 825 (2002).

\bibitem{fried58}
{J.~Friedel}, Nuovo Cim.~Suppl. {\bf {7}}, 287 (1958).

\bibitem{yang92}
{S.-R.~E.~Yang}, Solid State Comm. {\bf {81}}, 375 (1992).

\bibitem{butti93}
{M.~B\"uttiker, A.~Pretre, and H.~Thomas}, Phys.~Rev.~Lett. {\bf {70}}, 4114
(1993); Phys.~Rev.~B {\bf 54}, 8130 (1996).

\bibitem{blant98}
{Ya.~M.~Blanter, F.~W.~J.~Hekking, and M.~B\"uttiker}, Phys.~Rev.~Lett. {\bf
  {81}}, 1925 (1998).

\bibitem{fabri95}
{M.~Fabrizio and A.~O.~Gogolin}, Phys.~Rev.~B {\bf {51}}, 17827 (1995). 

\bibitem{eggra97}
{R.~Egger and H.~Grabert}, Phys.~Rev.~Lett. {\bf {79}}, 3463
(1997).

\bibitem{maslov}
{D.L.~Maslov and M.~Stone}, Phys.~Rev.~B {\bf {52}}, R5539 (1995);
{V.V.~Ponomarenko}, {\em {ibid.}} {\bf {52}}, R8666 (1995); {I.~Safi and
H.J.~Schulz}, {\em {ibid.}} {\bf {52}}, R17040 (1995).

\end{thebibliography}
\end{document}